\documentclass[pdflatex,sn-nature]{sn-jnl}

\usepackage{xspace}
\usepackage{graphicx}%
\usepackage{multirow}%
\usepackage{amsmath,amssymb,amsfonts}%
\usepackage{amsthm}%
\usepackage{mathrsfs}%
\usepackage[title]{appendix}%
\usepackage{xcolor}%
\usepackage{textcomp}%
\usepackage{manyfoot}%
\usepackage{booktabs}%
\usepackage{algorithm}%
\usepackage{algorithmicx}%
\usepackage{algpseudocode}%
\usepackage{listings}%

\newcommand{\txs}{TXS\,0506+056\xspace}
\newcommand{\gb}{GB6\,J1542+6129\xspace} 
\newcommand{\cgcg}{CGCG\,420-015\xspace} 
\newcommand{\pks}{PKS\,1424+240\xspace}
\newcommand{\nustar}{\textit{NuSTAR}\xspace}

\newcommand{\flux}{\,erg\,s$^{-1}$\,cm$^{-2}$\xspace}

\newcommand*\aap{A\&A}

\newcommand*\aj{AJ}

\newcommand*\apjl{ApJL}
\newcommand*\apj{ApJ}

\newcommand*\apjs{ApJS}

\newcommand*\jcap{J. Cosmology Astropart. Phys.}

\newcommand*\mnras{MNRAS}

\newcommand*\prd{Phys.~Rev. D}
\newcommand*\prl{Phys.~Rev. Lett.}

\theoremstyle{thmstyleone}%
\theoremstyle{thmstyletwo}%

\theoremstyle{thmstylethree}%

\begin{document}

\title{A Seyfert galaxy as a hidden counterpart to a neutrino-associated blazar}

\author*[1,2,3,4,5]{\fnm{Emma} \sur{Kun}}\email{kun.emma@csfk.org}

\author[6]{\fnm{Santiago} \sur{del Palacio}}

\author[7]{\fnm{Imre} \sur{Bartos}}

\author[8]{\fnm{Francis} \sur{Halzen}}

\author[1,3,6]{\fnm{Julia Becker} \sur{Tjus}}

\author[9,10]{\fnm{Peter L.} \sur{Biermann}}

\author[2]{\fnm{Anna} \sur{Franckowiak}}

\author[11,12]{\fnm{Claudio} \sur{Ricci}}


\affil*[1]{\orgdiv{Theoretical Physics IV: Plasma-Astroparticle Physics, Faculty for Physics \& Astronomy}, 
\orgname{Ruhr University Bochum}, \orgaddress{\city{Bochum}, \postcode{44780}, \country{Germany}}}

\affil[2]{\orgdiv{Astronomical Institute, Faculty for Physics \& Astronomy}, 
\orgname{Ruhr University Bochum}, \orgaddress{\city{Bochum}, \postcode{44780}, \country{Germany}}}

\affil[3]{\orgdiv{Ruhr Astroparticle and Plasma Physics Center (RAPP Center)}, 
\orgname{Ruhr University Bochum}, \orgaddress{\city{Bochum}, \postcode{44780}, \country{Germany}}}

\affil[4]{\orgname{Konkoly Observatory, HUN-REN Research Centre for Astronomy and Earth Sciences}, 
\orgaddress{\street{Konkoly Thege Miklós út 15-17}, \city{Budapest}, \postcode{H-1121}, \country{Hungary}}}

\affil[5]{\orgname{CSFK, MTA Centre of Excellence}, 
\orgaddress{\street{Konkoly Thege Miklós út 15-17}, \city{Budapest}, \postcode{H-1121}, \country{Hungary}}}

\affil[6]{\orgdiv{Department of Space, Earth and Environment}, 
\orgname{Chalmers University of Technology}, \orgaddress{\city{Gothenburg}, \postcode{SE-412 96}, \country{Sweden}}}

\affil[7]{\orgdiv{Department of Physics}, \orgname{University of Florida}, 
\orgaddress{\street{PO Box 118440}, \city{Gainesville}, \state{FL}, \postcode{32611-8440}, \country{USA}}}

\affil[8]{\orgdiv{Department of Physics}, \orgname{University of Wisconsin}, 
\orgaddress{\city{Madison}, \state{WI}, \postcode{53706}, \country{USA}}}

\affil[9]{\orgname{Max Planck Institute for Radio Astronomy}, 
\orgaddress{\city{Bonn}, \postcode{53121}, \country{Germany}}}

\affil[10]{\orgdiv{Department of Physics \& Astronomy}, \orgname{University of Alabama}, 
\orgaddress{\city{Tuscaloosa}, \state{AL}, \postcode{35487}, \country{USA}}}

\affil[11]{\orgdiv{Department of Astronomy}, \orgname{University of Geneva}, 
\orgaddress{\country{Switzerland}}}

\affil[12]{\orgdiv{Instituto de Estudios Astrofísicos, Facultad de Ingeniería y Ciencias}, 
\orgname{Universidad Diego Portales}, \orgaddress{\street{Av. Ejército Libertador 441}, \city{Santiago}, \country{Chile}}}

\abstract{
The origin and production mechanisms of high-energy astrophysical neutrinos remain open questions in multimessenger astronomy. Previous studies have hinted at a possible linear correlation between the hard X-ray and high-energy neutrino emission in active galactic nuclei. New \nustar observations, first presented here, reveal that blazar PKS~1424+240,  located within a prominent IceCube neutrino hotspot, is far fainter in hard X-rays than expected from this trend. Motivated by this apparent ambiguity, we identify the nearby Seyfert galaxy NGC~5610, also coincident with the hotspot, whose unabsorbed hard X-ray flux exceeds that of PKS~1424+240 by about an order of magnitude. When the local IceCube neutrino flux is apportioned between the two AGN in proportion to their hard X-ray emission, both align with the previously suggested X-ray–neutrino correlation. This suggests that certain IceCube hotspots may be unresolved blends of multiple AGN, and supports a multimessenger scenario in which high-energy neutrinos and hard X-rays originate from the same hadronic interactions, with the X-ray emission produced through cascade reprocessing.}




\maketitle

\section{Introduction}

Since the discovery of a diffuse TeV–PeV neutrino flux by the IceCube Neutrino Observatory \citep{2013Sci...342E...1I}, the nature of the astrophysical environments capable of producing such high-energy particles has remained uncertain. Active galactic nuclei (AGN) are among the leading candidates \citep{ICTXS2018a,2018Sci...361..147I,ngc1068_2022}, motivated by the presence of powerful accelerators near accreting supermassive black holes and by the dense ultraviolet and X-ray photon fields that enable efficient photohadronic ($p\gamma$) interactions. Recent IceCube observations and theoretical work suggest that the magnetized, turbulent coronae of nearby X-ray–bright Seyfert galaxies may be particularly effective at generating $\sim$1–10\,TeV neutrinos through $p\gamma$ interactions \citep{2025A&A...697A.124L}.

Blazar-type AGN, by contrast, are typically associated with jet-related neutrino production. Their relativistic jets, aligned close to our line of sight and generally optically thin to gamma rays, allow high-energy photons to escape without being reprocessed into the MeV or hard X-ray band. High-frequency–peaked BL Lac objects in particular lack strong external ultraviolet photon fields; their soft radiation is dominated by synchrotron emission and only weakly influenced by a broad-line region or dusty torus. As a result, photohadronic interaction efficiencies and internal $\gamma\gamma$ absorption are both low, limiting the production of hard X-rays from cascades \citep{Tavecchio2019, Murase2014, Zacharias2021}.

Despite the distinct environments of Seyferts and blazars, recent observations reveal a nearly linear correlation between the 0.3–100\,TeV IceCube neutrino luminosity and the 15–55\,keV \nustar hard X-ray luminosity across a small, but diverse group of AGN \citep{Kun2024}. Similar trends had been proposed for obscured AGN cores \citep[e.g.][]{2020PhRvL.125a1101M,2022ApJ...941L..17M,2021ApJ...911L..18K,Kun2023,Neronov2024}, but the emergence of a tight relation spanning four radio-quiet neutrino source-candidate Seyferts, NGC~1068 (global significance $4.2\sigma$ \citep{ngc1068_2022}), NGC~3079 ($TS \sim 14.1$ \citep{Neronov2024}), NGC~4151 ($3.2\sigma$ local
pretrial significance with power-law spectrum \citep{Abassi2024}), \cgcg ($3.5\sigma$ local
pretrial significance with a disk-corona model \citep{Abassi2024}), and two radio-loud neutrino source-candidate blazars, \txs (post-trial significance $3.5\sigma$ \citep{2018Sci...361..147I}), \gb (local pre-trial significance $2.9 \sigma$ \citep{IC2020tenyears}) was previously unrecognized. This correlation suggests that, in both classes, neutrino production may be powered by a continuous central engine rather than by transient jet activity. In such a framework, high-energy neutrinos originate near the black hole through $p\gamma$ interactions within dense ultraviolet photon fields associated with the inner accretion disk. The accompanying pionic gamma rays can undergo $\gamma\gamma \rightarrow e^{+}e^{-}$ absorption and initiate electromagnetic cascades that reprocess their energy into MeV and hard X-ray photons, producing the characteristic multimessenger signature of neutrino-bright, gamma-ray–dim, hard X-ray–luminous AGN cores. In blazars, where external photon fields are weaker, this neutrino–X-ray coupling may be suppressed, although neutrino-associated blazars appear unusually X-ray–luminous compared with the broader VLBI blazar population \citep{2024JCAP...05..133P}.

PKS~1424+240 is one of the most promising neutrino-associated radio-loud AGN in the Northern sky \citep{ngc1068_2022}, identified with a pre-trial significance of $\sim3.7\sigma$ in IceCube’s 10-year time-integrated analysis \citep{ngc1068_2022}. It is a high-frequency–peaked BL Lac object at $z\approx0.605$ and among the most distant persistently VHE-emitting ($E>100$\,GeV) blazars known \citep{magicpks2014}. PKS~1424+240 has been described as a “masquerading BL Lac” \citep{Padovani2022}, potentially hosting a hidden broad-line region and a radiatively efficient accretion disk, similar to TXS~0506+056. In such systems, neutrinos may originate either in the jet or in the corona, with secondary gamma rays partially attenuated by internal absorption. Broadband SED modelling confirms that hadronic and hybrid scenarios can reproduce its VHE emission without exceeding the Eddington limit \citep{Cerruti2015}. These physical ambiguities make PKS~1424+240 a key test case for assessing whether the proposed hard X-ray–neutrino connection extends to blazars, motivating the hard X-ray observations presented in this work.

To assess this possibility, we obtained hard X-ray observations of PKS~1424+240 with \nustar. As we report below, the blazar is substantially fainter in the 15–55\,keV band than expected from the proposed X-ray–neutrino trend, a discrepancy that calls for an explanation. One possibility is that other sources in the region contribute to the local neutrino emission. Interestingly, we find that the Seyfert galaxy NGC~5610 lies only about $1^\circ$ from PKS~1424+240. NGC~5610 (fiducial position in wide-band IR with \textit{WISE}: $\mathrm{RA}_{J2000}=216.0956^\circ$, $\mathrm{DEC}_{J2000}=24.6142^\circ$ \citep{2013wise}, $z=0.016922$) hosts an AGN with a faint radio core, exhibiting a 1.4\,GHz flux density of 31.4\,mJy \citep{condon2002}. NGC~5610 is not detected in gamma rays, and was not previously been considered in IceCube association analyses with Seyfert galaxies \citep{Abassi2024}.

In this work, we investigate the region around PKS~1424+240 with the aim of disentangling potential contributors to the local neutrino signal. By distributing the IceCube neutrino excess between PKS~1424+240 and NGC~5610 in proportion to their hard X-ray fluxes, we reassess and extend the proposed hard X-ray–neutrino luminosity correlation originally observed in four Seyferts and two blazars \citep{Kun2024}. This approach allows us to test whether the unexpectedly low hard X-ray emission of PKS~1424+240 can be understood within a composite-source scenario and to evaluate the robustness of the correlation when multiple AGN contribute to the same IceCube hotspot.

\section{Results}

We observed \pks with \nustar \citep{NuSTAR2013} under program 10049 (PI del Palacio) for 27\,ks on February 4, 2025. Details on the data analysis are given in Sect.~\ref{sec:methods}. The observed spectrum is well described by a power-law with a 15–55\,keV flux of  $F_\mathrm{hX}^\mathrm{PKS} = (4.9\pm1.0)\times10^{-13}$\flux in the framework of the observer and a photon index $\Gamma = 1.70 \pm 0.14$. The source might be in a low X-ray state: its 2–10 keV flux is a factor of $\sim 3$ below typical Swift X-Ray Telescope \citep[Swift-XRT,][]{2017ApJS..233....3T} levels since MJD 58000 (excluding earlier flares, see e.g. \citep{pks1424-2023}). The observed hardening of the spectrum (compared to the Swift-XRT soft-band index, $\Gamma \sim 2.4$) resembles that seen in other neutrino-associated AGN such as TXS 0506+056 \citep{ICTXS2018a}; however, we note that the physical mechanisms behind the spectral shapes differ across AGN classes, with blazar spectra shaped by jet-related processes and Seyfert spectra influenced by absorption and reprocessing, as in the case of NGC 1068 \citep{ngc1068_2022}.

In IceCube’s 10-year Pass2 track dataset \citep{ngc1068_2022_dataset}, which contains uniformly processed full detector data, PKS~1424+240 is associated with best-fit neutrino number $\hat{n}_s = 77$ and astrophysical spectral index $\hat{\gamma} = 3.5$, corresponding to a pre-trial significance of $3.7\sigma$ ($\approx 2.3 \sigma$ post-trial in a 110 source-catalog search). We note that the IceCube flux estimates are maximum-likelihood estimate (MLE) values rather than upper limits. Nevertheless, the absence of a $>3\sigma$ post-trial detection places a practical constraint on the allowed neutrino luminosity: a substantially higher true flux would be statistically disfavoured, as it would have produced a more significant excess in the 110-source catalog search. The inferred, somewhat limiting 0.3–100\,TeV muon-neutrino flux is $F_{\nu} \approx 5.1\times10^{-11}$\flux considering the above best-fit values based on the full detector IceCube dataset, collected between 2011 and 2020. Combining this value with our \nustar measurement places PKS~1424+240 far from the established hard X-ray–neutrino correlation, since the neutrino flux is nearly two orders of magnitude higher than expected given its hard X-ray emission. This prompted us to revisit our interpretation.

Radio interferometric monitoring with the Very Long Baseline Array (VLBA) at 15~GHz shows that the Doppler factor of the VLBI core of PKS~1424+240 increased by a factor of $\sim 8$ after 2014, which remained elevated until 2019–2021 \citep{pks1424-2023}. This interval overlaps with a strong soft X-ray brightening in \textit{Swift}-XRT data (2016–2018), suggesting jet activity that may not be reflected in the hard X-ray band. IceCube’s multiflare analysis indicates a neutrino flare of approximately 200 days in duration, peaking on June 9, 2015, with modest local pre-trial significance ($p = 0.1$, \citep{icecubemultiflare}). However, the absence of corresponding \textit{Fermi}-LAT variability and the lack of hard X-ray coverage leave the jet scenario inconclusive. 

\begin{figure*}
    \centering
     \includegraphics[width=0.65\linewidth]{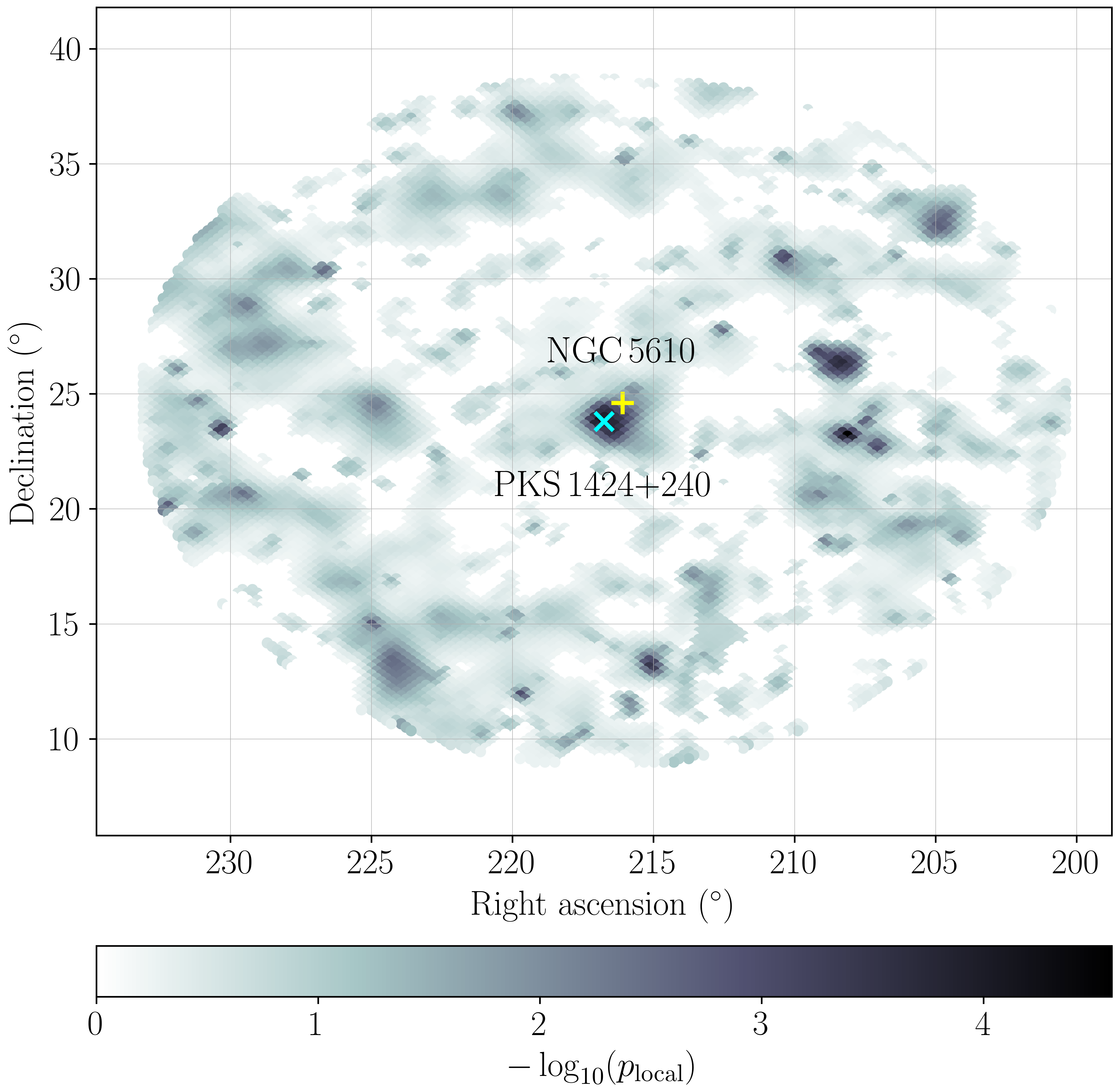}
    \caption{Reconstructed sky map of the scan for point sources in a 15 degrees radius about PKS 1424+240 with free spectral index, using Hammer-Aitoff projection (seasons IC86 I-II, full detector data between 2011 and 2020; \citep{ngc1068_2022_dataset}). The fiducial sky positions of PKS~1424+240 and NGC~5610 are marked with a light blue cross and a yellow plus, respectively.}
    \label{fig:skymap-zoom}
\end{figure*}

A more compelling explanation arises after realizing the presence of the Seyfert galaxy NGC~5610, located only about $1^\circ$ from PKS~1424+240. In Fig. \ref{fig:skymap-zoom}, we show the reconstructed sky map of the maximum likelihood scan for point sources in a 15 degrees radius about PKS 1424+240 with free spectral index, using Hammer-Aitoff projection (seasons IC86 I-II, full detector data between 2011 and 2020; \citep{ngc1068_2022_dataset}). Although the local $p$-value minimum in the IceCube skymap appears closer to PKS 1424+240, both it and NGC 5610 lie within IceCube’s angular resolution, allowing potential contributions from either or both sources. We note that while IceCube's per-event angular resolution is $\sim 1^\circ$, the localization uncertainty of a hotspot with $n_s \approx 77$ is smaller, $\sim 0.11^\circ$, making PKS 1424+240 a more probable positional match than NGC 5610. However, the soft spectral index, $\gamma = 3.5$, and lack of strong localization preference in the skymap (Fig. \ref{fig:skymap-zoom}) allow the possibility of a composite origin, particularly given the proximity and high hard X-ray flux of NGC 5610.

The H column density for NGC~5610 is estimated to be $N_\mathrm{H} = 3.63^{+0.54}_{-0.75}\times 10^{22}\,\mathrm{cm}^{-2}$ \citep{Ricci2017}, while the Galactic H column density reported is $N_\mathrm{H,Gal} \approx 1.9 \times 10^{20}\,\mathrm{cm}^{-2}$ \citep{2005A&A...440..775K}. From its \textit{Swift}/BAT 0.3–150\,keV flux of $1.92\times10^{-11}$\flux and photon index $\Gamma = 1.58 \pm 0.23$ \citep{Ricci2017}, we derive an unabsorbed 15–55\,keV flux $F_\mathrm{hX}^\mathrm{NGC} = (4.72 \pm 0.43)\times10^{-12}$\flux assuming the above column density $N_\mathrm{H}$. The derived 15--55 keV hard X-ray flux of NGC 5610 is thus more than an order of magnitude brighter than that of PKS~1424+240.

Given IceCube’s angular resolution, we consider a dual-source interpretation in which the total neutrino excess $F_{\nu}^\mathrm{tot} \approx 5.1\times10^{-11}$\flux between 0.3 and 100 TeV arises from both AGN in proportion to their hard X-ray fluxes. Using $F_\mathrm{hX}^\mathrm{PKS} = 4.9\times10^{-13}$\flux and $F_\mathrm{hX}^\mathrm{NGC} = 4.7\times10^{-12}$\flux, the ratio of them yields the contributions from the two AGN to the total neutrino excess as
\[
F_{\nu}^\mathrm{PKS} \approx 4.8\times10^{-12}\ \mathrm{erg\,s^{-1}\,cm^{-2}}, \qquad
F_{\nu}^\mathrm{NGC} \approx 4.6\times10^{-11}\ \mathrm{erg\,s^{-1}\,cm^{-2}}.
\]

\begin{figure*}
    \centering    
    \includegraphics[width=0.75\textwidth]{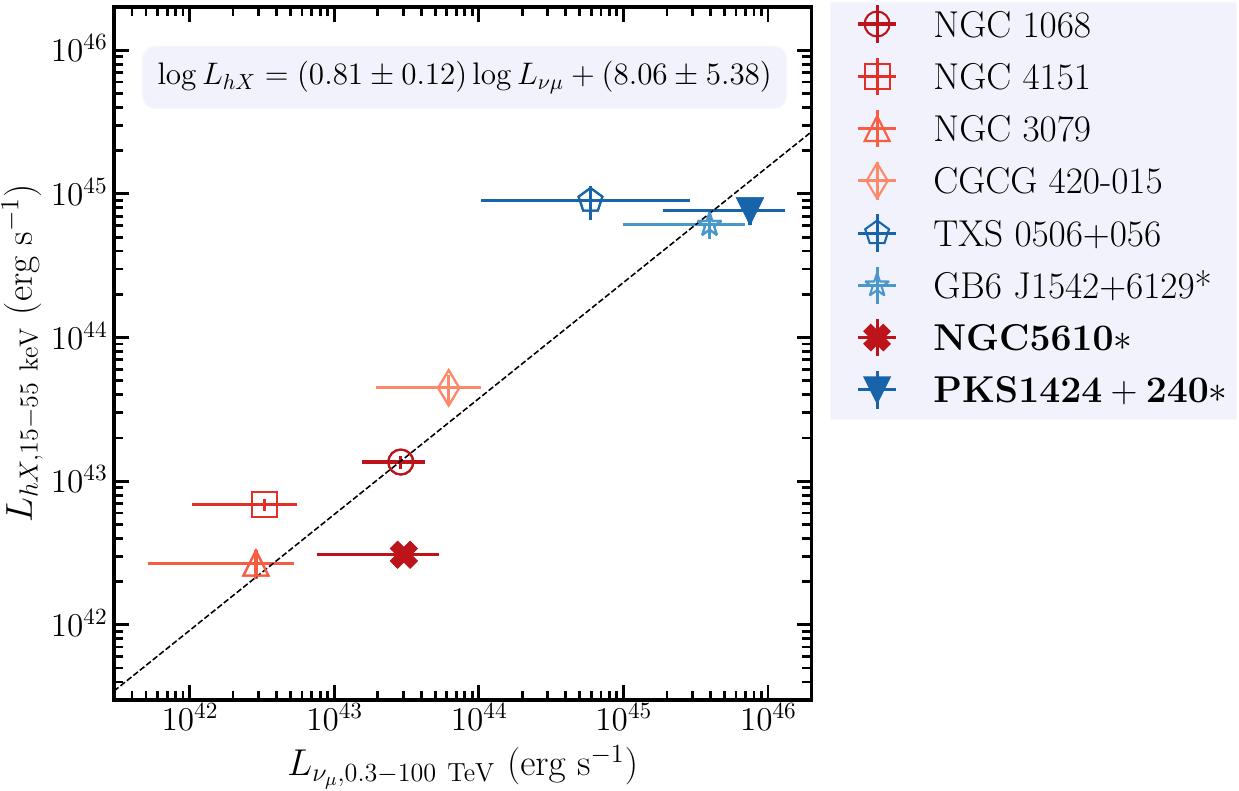}
    \caption{Correlation between unabsorbed hard X-ray and neutrino luminosities for radio-quiet Seyferts (red) and radio-loud blazars (blue). NGC~5610 and PKS~1424+240 are shown with updated luminosities under the dual-source scenario. These two additional sources are highlighted with filled markers and boldface in the legend. The dashed line corresponds to the fitted line, whose equation is shown on the top.  For sources marked with $^*$, the neutrino luminosity comes from limiting IceCube MLE fluxes.}
    \label{fig:corrplot-extended}
\end{figure*}

We show in Fig.~\ref{fig:corrplot-extended} the hard X-ray–neutrino correlation plot from \citep{Kun2024}, now extended the 0.3–100 TeV neutrino and 15–55 keV hard X-ray luminosities of NGC 5610 and PKS 1424+240. These luminosities are derived from the above fluxes assuming a $\Lambda$CDM cosmology with $H_0=69.6~\mathrm{km,s^{-1},Mpc^{-1}}$, $\Omega_{\mathrm{m},0}=0.286$, $\Omega_{\lambda,0}=0.714$, and $T_{\mathrm{cmb},0}=2.72548~\mathrm{K}$. The two new sources follow the previously established correlation, as discussed in the next section. For the sources without significant IceCube detections ($<3\sigma$ post-trial), we adopt the IceCube MLE neutrino fluxes and assign a conservative 75\% fractional uncertainty to reflect the broad likelihood width at low significance.

\section{Discussion}

Our analysis extends the proposed connection between the high-energy neutrino and unabsorbed hard X-ray luminosity in AGN \citep{Kun2024} by adding two new sources: the blazar PKS~1424+240 and the nearby Seyfert galaxy NGC~5610. To quantify how the expanded sample affects the correlation, we fitted the hard X-ray–neutrino luminosity relation in log–log space using orthogonal distance regression (Fig. \ref{fig:corrplot-extended}). The best-fit slope is $\alpha\approx0.81\pm0.12$, roughly consistent with a linear relation. 

In the following, we assess how the strength of the correlation changes when these two objects are incorporated. For the hard X-ray–neutrino luminosity relation reported for two blazars and four Seyferts \citep{Kun2024}, the Pearson coefficient emerged as $R_{P}=0.97$, indicating a strong linear trend. Because luminosity–luminosity correlations can be affected by redshift evolution, we computed partial correlation coefficients that control for redshift. The partial Spearman coefficient is $R_{pS}=0.852$ and the partial Pearson coefficient is $R_{pP}=0.737$, demonstrating that the correlation persists even after removing redshift-driven effects. We note that the apparent X-ray--neutrino luminosity correlation may be influenced by selection effects arising from flux thresholds in both neutrino and X-ray observations across a redshift-spanning source population. This can induce apparent luminosity correlations from intrinsically uncorrelated fluxes, particularly in the presence of steep logN–logS distributions, as discussed in the context of $IR–\gamma$ correlations \citep{2012ApJ...755..164A}. We mitigated this by computing partial correlation coefficients controlling for redshift, which remain significant as discussed above, but a full assessment of selection biases will require a larger, statistically complete sample.

When we include PKS~1424+240 and NGC~5610 in the composite-source scenario and update the luminosities accordingly, the Pearson coefficient becomes $R_{P}=0.919$ (partial: $R_{pP}=0.514$) and the Spearman coefficient as $R_{S}=0.857$ (partial: $R_{pS}=0.686$). These values show that the correlation remains strong when the total ten-year IceCube neutrino flux near PKS 1424+240 is jointly attributed to PKS~1424+240 and NGC~5610 in proportion to their 15--55 keV hard X-ray emission. This correlation likely reflects efficient proton–photon interactions in compact regions in AGN cores, where high-energy neutrinos and hard X-rays are both produced through hadronic cascades. In this framework, the hard X-ray luminosity serves as a tracer of target photon fields with high enough $p\gamma$ optical depth to enable efficient neutrino production.

By redistributing the total IceCube-reported neutrino flux according to the hard X-ray flux ratio between the two AGN, we find that both PKS~1424+240 and NGC~5610 align with the observed correlation. This dual-source interpretation naturally explains the elevated neutrino flux in this region and highlights the possibility that some IceCube point sources may represent unresolved blends of multiple AGN within the current angular resolution. Similar source confusion could arise in other IceCube hotspots, especially in regions with multiple X-ray–bright AGN within $\sim 0.1^\circ-2^\circ$ comparable to IceCube’s angular uncertainty with track neutrino events.

To summarize, our main findings are:
\begin{itemize}
\item We identified NGC~5610 as a plausible, previously unrecognized contributor to the local IceCube neutrino excess near PKS 1424+240.
 \item We extended the previously reported hard X-ray–neutrino luminosity sample from six \citep{Kun2024} to eight AGN, further strengthening the correlation, now spanning three blazar- and five Seyfert-type active galaxies.
\item We demonstrate that indirect identification of neutrino sources is achievable through multimessenger inference.
\end{itemize}

More broadly, our results support a picture in which high-energy neutrinos from AGN arise predominantly from $p\gamma$ interactions in dense UV/X-ray photon fields near the central black hole, such as those found in hot coronae. This framework naturally accommodates both radio-quiet Seyferts and radio-loud AGN, and explains why some neutrino-emitting blazars appear X-ray–bright compared with the general blazar population. However, in jet-dominated systems, neutrino production may proceed independently of the hard X-ray output because of the scarcity of target photons in the jet environment, leading to neutrino sources faint in hard X-rays.

While the observed correlation between hard X-ray and neutrino luminosities suggests a physical link, it does not imply that all X-ray–bright AGN are efficient neutrino emitters. The trend is expected to trace environments where photohadronic interactions are efficient,  typically requiring compact regions with dense photon fields and sustained proton acceleration, such as coronae near accreting supermassive black holes. X-ray luminosity alone is therefore a necessary but not sufficient signature of efficient neutrino production. AGN that are bright in hard X-rays but undetected in neutrinos may lack the specific physical conditions (e.g., target photon compactness, high-energy protons, or sufficient baryon loading) required for significant $p\gamma$ neutrino output. As such, the correlation should be interpreted as a diagnostic of hadronic interaction efficiency, high $p\gamma$ optical depths, rather than as a universal predictor of neutrino emission.

Several caveats remain. One is that differences in observational windows and non-simultaneous data can introduce scatter into the correlation, particularly when the timescales of neutrino and X-ray observations are not matched. For example, the neutrino hotspot on the $p$-value map (see Fig. \ref{fig:skymap-zoom}) seems to peak closer to PKS 1424+240. Given the variability of blazar jets, PKS 1424+240 may have contributed during the 2011–2020 IceCube window through episodic jet activity through a transient $p \gamma$ event, even though it appears faint in our 2025 NuSTAR observation. The relative contribution of $pp$ interactions is also uncertain. Deviations up to an $\sim$order of magnitude along either axis are therefore not unexpected. Nonetheless, the consistency of multiple sources across different AGN classes suggests a robust underlying physical connection. We also note that attributing neutrino flux in proportion to X-ray flux and subsequently verifying a luminosity correlation may introduce an element of circular reasoning. However, since the two AGN lie at different redshifts, the resulting luminosities do not scale linearly with the partitioning assumption, preserving some independence.

Taken together, the hard X-ray-neutrino luminosity relation provides a physically motivated pathway for identifying AGN neutrino source candidates. At least until hard X-ray instruments improve in sensitivity and future neutrino observatories deliver better angular resolution and statistics, this multimessenger approach may prove essential for revealing the high-energy neutrino sky and the environments capable of producing the most energetic particles in the Universe.

\section{Methods}\label{sec:methods}

\begin{figure}[h!]
    \centering
    \includegraphics[width=0.63\linewidth, trim=0cm 0.2cm 3.9cm 0cm, clip]{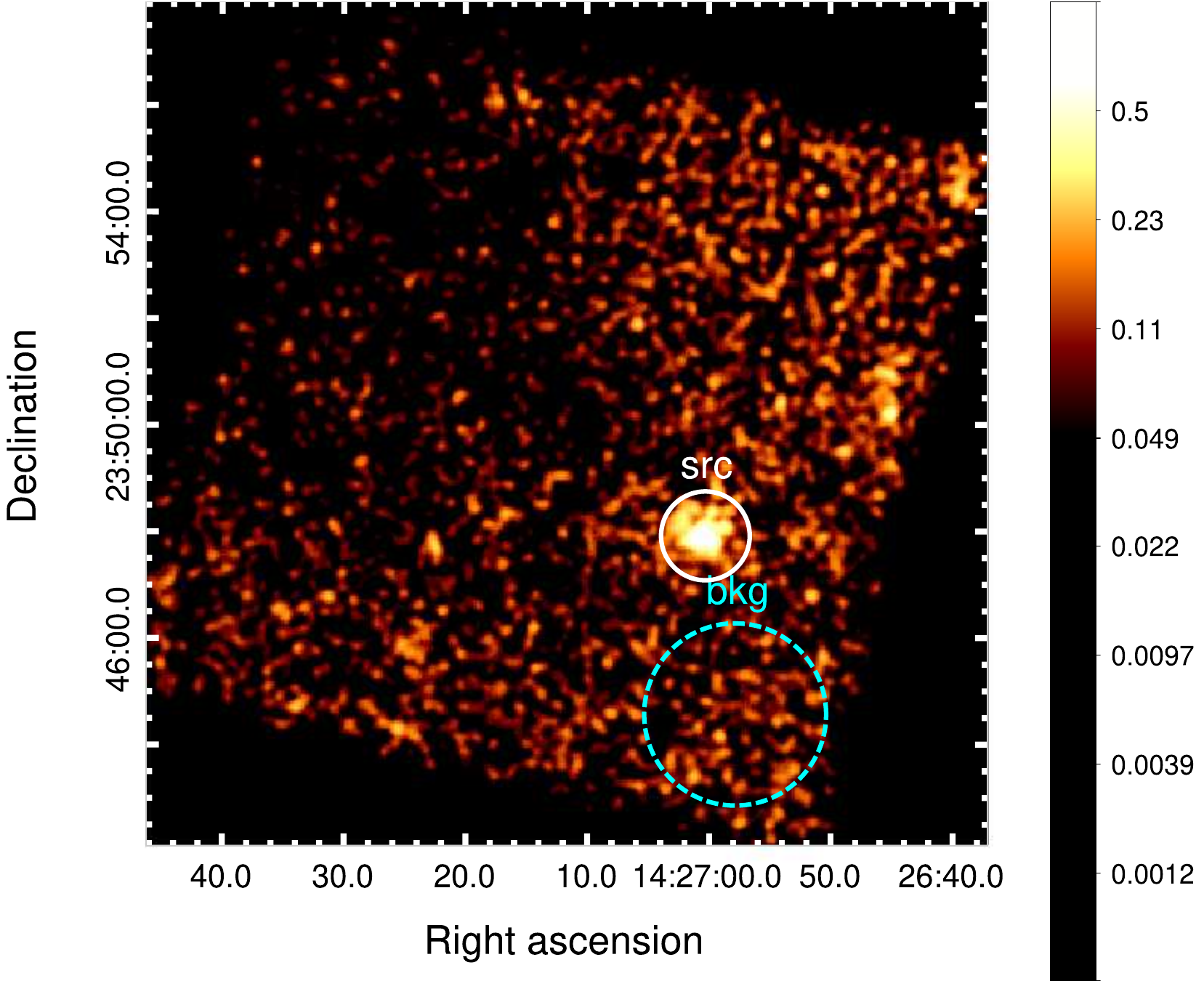}
    \caption{\textit{NuSTAR}-FPMB intensity image in the 3--20\,keV band. Source and background extraction regions are shown.}
    \label{fig:nustar_pks1424-field}
\end{figure}

We reduced the \nustar data using \texttt{Heasoft v.6.30.1} and the task \textit{nupipeline} with the flags \textit{saacalc=2 saamode=optimized tentacle=yes} to remove high-background times (which affected $<$1\% of the data). 
We used the task \textit{nuproducts} to extract the source spectrum within a 50" aperture, and the background spectrum in a large region devoid of point sources within the same chip. In Fig.~\ref{fig:nustar_pks1424-field} we show the field-of-view and the regions used in the analysis.

\begin{figure}[h!]
    \centering
    \includegraphics[width=0.8\linewidth]{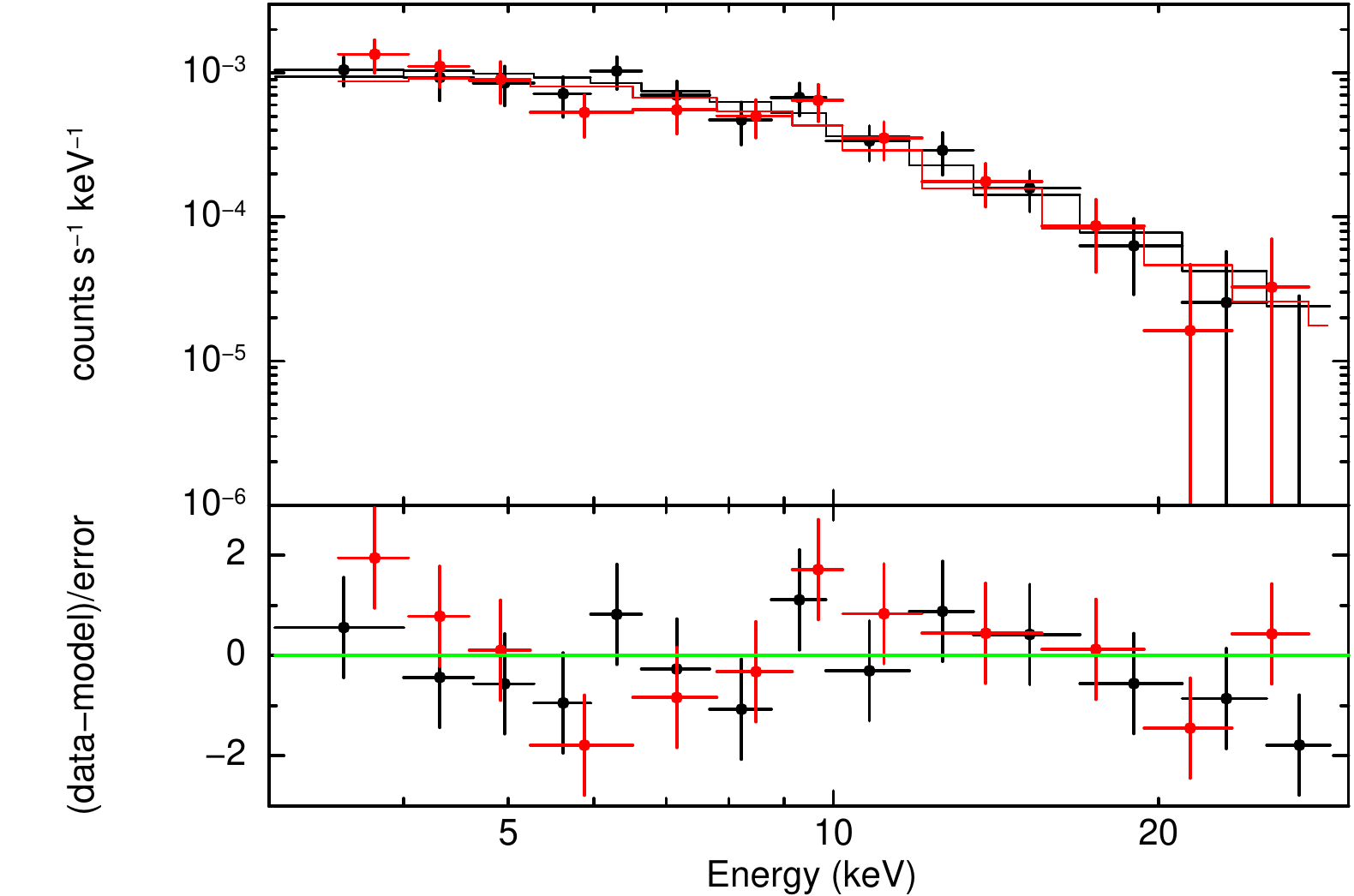}
    \caption{\nustar X-ray spectrum of \pks (top panel) and residuals of the fit (bottom panel). Black and red symbols correspond to FPMA and FPMB, respectively. The fitted model is a power law.}
    \label{fig:nustar_spectrum}
\end{figure}

The spectra were rebinned with the task \textit{ftgrouppha} and the flag \textit{grouptype=opt}.
The source is detected above the background up to $\sim 20$ keV. The spectrum was fitted with a power-law model in \texttt{XSPEC v.12.12.1} \citep{Arnaud1996}, using Cash statistics to account for the low number of counts at high energies. The spectral fit is very good (C-stat = 114.64/116\,d.o.f.), and it yields a flux of $F_\mathrm{hX}^\mathrm{PKS} = (4.9\pm1.0)\times10^{-13}$\flux and a photon index of $\Gamma=1.70\pm0.14$. We show the spectrum in Fig.~\ref{fig:nustar_spectrum}.

\bmhead{Acknowledgements}

The authors thank Ali Kheirandish, Yuri Kovalev, and Justin Vandenbroucke for fruitful discussions and their helpful comments that improved the quality of the paper. E.K. thanks the Bundesministerium für Forschung, Technologie und Raumfahrt. J.B.T.\, E.K., and A.F.\ acknowledge support from the German Science Foundation DFG, via the Collaborative Research Center \textit{SFB1491: Cosmic Interacting Matters -- from Source to Signal} (grant no.\ 445052434). S.d.P. acknowledges support from ERC Advanced Grant 789410. CR acknowledges support from SNSF Consolidator grant F01$-$13252, Fondecyt Regular grant 1230345, ANID BASAL project FB210003 and the China-Chile joint research fund. This research has made use of the \nustar Data Analysis Software (NuSTARDAS) jointly developed by the ASI Space Science Data Center (SSDC, Italy) and the California Institute of Technology (Caltech, USA). This work made use of Astropy: \url{http://www.astropy.org}, a community-developed core Python package and an ecosystem of tools and resources for astronomy \cite{astropy:2013, astropy:2022}.

\section*{Declarations}

\textbf{Conflict of interest/Competing interests} The authors declare no competing interests.\\
\textbf{Ethics approval and consent to participate} All authors approved the manuscript and consented to participate in the work without ethical issues.\\
\textbf{Author contribution:} E.K. prepared the initial manuscript draft. S.d.P. carried out the \textit{NuSTAR} data analysis. All authors contributed to the interpretation of the results and the final shaping of the manuscript.\\
\textbf{Data availability}  
All data used in this study are publicly available. Results based on IceCube Pass2 data are accessible through the Collaboration paper \citep{ngc1068_2022_dataset}, and \textit{NuSTAR} data can be downloaded under ObsID 10049.\\
\textbf{Code availability} 
Custom analysis scripts that use fully public data, are available from the corresponding author upon reasonable request.\\
\textbf{Consent for publication}
Consent for publication is given by all authors.

\end{document}